\documentclass[twocolumn]{aastex631}

\usepackage{amsthm}
\usepackage{amsmath}
\usepackage{hyperref}
\usepackage{booktabs}
\usepackage{float}
\usepackage{url}
\usepackage{grffile}
\usepackage{gensymb}
\usepackage{bigints}

\graphicspath{{./} {../figures/}}

\newcommand{\Rom}{\textsc{Romulus25}}
\newcommand{\HI}{\hbox{\rmfamily H\,{\scshape i}\;}}
\newcommand{\ergs}{erg\,s$^{-1}$\;}

\begin{document}

\title{A hidden population of massive black holes in simulated dwarf galaxies}
\shorttitle{Hidden MBHs in \Rom{}}
\shortauthors{Sharma et al.}

\author{Ray S. Sharma}
\affiliation{Department of Physics and Astronomy, Rutgers, The State University of New Jersey, 136 Frelinghuysen Road, Piscataway, NJ 08854, USA}

\author{Alyson M. Brooks}
\affiliation{Department of Physics and Astronomy, Rutgers, The State University of New Jersey, 136 Frelinghuysen Road, Piscataway, NJ 08854, USA}
\affiliation{Center for Computational Astrophysics, Flatiron Institute, 162 Fifth Avenue, New York, NY 10010, USA}

\author{Michael Tremmel}
\affiliation{Department of Astronomy, Yale University, 52 Hillhouse Ave, New Haven, CT 06511, USA}

\author{Jillian Bellovary}
\affiliation{Department of Physics, Queensborough Community College, City University of New York, 222--05 56th Ave, Bayside, NY 11364, USA}
\affiliation{Department of Astrophysics, American Museum of Natural History, Central Park West at 79th Street, New York, NY 10024, USA}
\affiliation{The Graduate Center, City University of New York, 365 5th Ave, New York, NY 10016, USA}

\author{Angelo Ricarte}
\affiliation{Center for Astrophysics $|$ Harvard \& Smithsonian, 60 Garden Street, Cambridge, MA 02138, USA}
\affiliation{Black Hole Initiative at Harvard University, 20 Garden Street, Cambridge, MA 02138, USA}

\author{Thomas R. Quinn}
\affiliation{Department of Astronomy, University of Washington, PO Box 351580, Seattle, WA 98195, USA}

\begin{abstract}
    We explore the characteristics of actively accreting MBHs within dwarf galaxies in the \Rom{} cosmological hydrodynamic simulation. We examine the MBH occupation fraction, x-ray active fractions, and AGN scaling relations within dwarf galaxies of stellar mass $10^{8} < M_{\rm star} < 10^{10} M_\odot$ out to redshift $z=2$. In the local universe, the MBH occupation fraction is consistent with observed constraints, dropping below unity at $M_{\rm star} < 3\times10^{10} M_{\odot}$, $M_{\rm 200} < 3\times10^{11} M_\odot$. Local dwarf AGN in \Rom{} follow observed scaling relations between AGN x-ray luminosity, stellar mass, and star formation rate, though they exhibit slightly higher active fractions and number densities than comparable x-ray observations. Since $z=2$, the MBH occupation fraction has decreased, the population of dwarf AGN has become overall less luminous, and as a result, the overall number density of dwarf AGN has diminished. We predict the existence of a large population of MBHs in the local universe with low x-ray luminosities and high contamination from x-ray binaries and the hot interstellar medium that are undetectable by current x-ray surveys. These hidden MBHs make up $76\%$ of all MBHs in local dwarf galaxies, and include many MBHs that are undermassive relative to their host galaxy's stellar mass. Their detection relies not only on greater instrument sensitivity but on better modeling of x-ray contaminants or multi-wavelength surveys. Our results indicate dwarf AGN were substantially more active in the past despite being low-luminosity today, and indicate future deep x-ray surveys may uncover many hidden MBHs in dwarf galaxies out to at least $z=2$.
\end{abstract}

\section{Introduction}

    Observations over the past two decades have begun to explore the prevalence of massive black holes (MBHs) in dwarf galaxies \citep{Shields2008, Reines2013, Moran2014, Miller2015, Trump2015, Nguyen2018, Kaviraj2019, Baldassare2020}. There are now many observations out to $z\sim2$ that indicate dwarf galaxies are capable of hosting actively accreting MBHs \citep{Reines2013,Lemons2015,Reines2015,Pardo2016,Baldassare2017,Ahn2018,Baldassare2018,Martin-Navarro2018,Mezcua2018,Birchall2020,Molina2021a}, albeit at lower luminosities than in massive galaxies \citep{Kormendy2013}. Recent studies have leveraged the observed active galactic nuclei (AGN) fraction into constraints on the occupation fraction of MBHs in dwarf galaxies \citep{Trump2015,Miller2015,Mezcua2018,Baldassare2020,Birchall2020}.
    
    Dwarf galaxies in the local universe provide a laboratory for studying how MBHs may have formed and grown in the early universe. Theoretical work suggests that MBHs in local dwarf galaxies have grown little relative to those found in massive galaxies \citep{Volonteri2008, Bellovary2019}. Hence MBHs in dwarf galaxies may provide insight into the early conditions of MBH seeding and growth, potentially explaining the origin of supermassive black holes found in the early universe \cite[e.g,][]{Vestergaard2009, Willott2010, Mortlock2011, Venemans2013, Wu2015}. Several mechanisms have been proposed
    which form either light seeds $\left(M_{\rm seed} \sim 10-10^{3} M_\odot\right)$ \citep{Devecchi2009,Davies2011,Whalen2012,Madau2014,Katz2015,Taylor2014,Yajima2016} or heavy seeds $\left(M_{\rm seed} \gtrsim 10^{4} M_\odot\right)$ \citep{Begelman2006,Johnson2013,Ferrara2014}. Each formation mechanism requires different density and metallicity characteristics of the local environment, and may ultimately imprint themselves on the MBH occupation fraction or mass function \citep{Lodato2006,Volonteri2009,Ricarte2018}. See \citet[][]{Volonteri2010,Volonteri2012,Latif2016,Greene2020} for in-depth reviews of MBH seeding mechanisms.
    
    MBHs within dwarf galaxies are difficult to detect outside the local universe because of their small sphere of influence (of order $0.1$ pc for an MBH of mass $M_{\rm BH} = 10^{5} M_\odot$). A small sphere of influence will 1) necessitate high resolving power in order to make dynamical MBH detections, and 2) restrict accretion rates, leading to less AGN activity and hence lower likelihood of detection \citep{Baldassare2016}. Recent searches for MBHs in dwarf galaxies have thus relied on searching for AGN via optical emission-line diagnostics \citep{Barth2004,Greene2004,Peterson2005,Greene2007,Reines2013,LaFranca2015,Sartori2015,Bentz2016,Bentz2016a,Marleau2017,Onori2017,Chilingarian2018}, IR color selection \citep{Satyapal2014}, IR/optical coronal line emission \citep{Satyapal2007,Satyapal2008,Satyapal2009,Cann2018,Cann2020,Cann2021,Satyapal2021,Molina2021,Bohn2021}, nuclear x-ray emission \citep{Mezcua2018, Birchall2020}, and optical variability \citep{Heinis2016,Baldassare2020}. See \citet{Mezcua2017} for an overview of the various techniques for observing dwarf AGN.
    
    Observations indicate AGN may play a role in dwarf galaxy evolution by impacting cold star-forming gas, reminiscent of the feedback found in massive galaxies \citep{Fabian2012, Kormendy2013, Somerville2015}. \citet{Bradford2018} connect AGN-like line ratios to \HI gas depletion and quiescence in $M_{\rm star} = 10^{9.2 - 9.5} M_\odot$ galaxies in the ALFALFA $70$\% survey \citep{Haynes2011}. \citet{Penny2018} identify ionized gas kinematically-offset from stars in $5$ SDSS-IV AGN with stellar masses $M_{\rm star} < 10^{9.3} M_\odot$. \citet{Manzano-King2019} find $13$ dwarf AGN with high velocity ionized gas outflows, where $6$ have outflows with AGN-like line ratios. \citet{Dickey2019} detect AGN-like hard-ionizing radiation in $16$ of $20$ low-mass, isolated, quiescent galaxies with stellar masses $M_{\rm star} = 10^{9.0 - 9.5} M_\odot$. \citet{Liu2020} identify  high-velocity, AGN-driven outflows in $8$ dwarf galaxies, where a small portion of outflowing material appear to escape into the circumgalactic medium.
    
    On the theory side, the question of AGN impact on dwarf galaxy evolution has yielded mixed results. Cosmological simulations have historically ignored AGN in dwarf galaxies due to resolution limitations, or assumptions on the inefficiency of AGN feedback in low-mass galaxies \cite[e.g,][]{Sijacki2015}. Only in recent years has it become more common to allow MBHs to form in galaxies below $M_{\rm star} \lesssim 10^{9.5} M_\odot$ within cosmological simulations \citep[e.g,][]{Habouzit2017}. Analytical models have found that dwarf AGN can eject gas with higher efficiencies than supernovae \citep{Dashyan2018}. Such a mechanism may help resolve certain anomalies in concordance $\Lambda$CDM cosmology \citep{Silk2017}. \citet{Koudmani2019} explore the impact of AGN feedback by applying various AGN activity models to a high-resolution isolated dwarf galaxy simulation. They find that dwarf AGN do not directly impact star formation but do drive hotter, faster outflows that may inhibit gas inflows. Using cosmological hydrodynamic simulations, \citet{Barai2019} detect early $z>4$ star formation suppression via AGN feedback. \citet{Sharma2020} find that \Rom{} dwarf galaxies that form relatively over-massive MBHs exhibit both suppressed star formation and depleted \HI gas. \citet{Koudmani2021} use the {\sc FABLE} cosmological simulation to study population statistics of AGN in dwarf galaxies. They find AGN can drive the kinematic misalignment between ionized gas and stars observed by \citet{Penny2018}.
    
    On the other hand, many cosmological hydrodynamic simulations indicate that strong supernova (SN) feedback will disperse gas in the central regions of dwarf galaxies and preemptively halt accretion onto the MBH \citep{Dubois2015,Bower2017,Angles-Alcazar2017,Habouzit2017,Trebitsch2018,Barausse2020}. The MBH can only begin growing again once the halo grows large enough to confine gas inflows. In particular, \citet{Habouzit2017} find that early, strong supernova feedback is an important ingredient to suppress MBH growth and match the observed $M_{\rm BH} - M_{\rm star}$ relation in dwarf galaxies. Their SN feedback model drives winds that sweep away star-forming gas, thereby cutting off early MBH accretion. They find that a weaker thermal SN feedback model allows the MBHs to constantly grow, likely growing too large at low redshift. \citet{Barausse2020} find that MBH growth regulated by SN feedback is important for matching the bolometric luminosity function at high redshift. SN feedback suppresses MBH growth and reduces the number of low-luminosity AGN, particularly in simulations with light seeding mechanisms.
    
    Recent analyses have shed light on the ability of cosmological simulations to model the low-mass MBHs found in low-mass galaxies. \citet{Haidar2022} perform a comparison of six large-scale $\left(>100\;{\rm cMpc}\right)^3$ cosmological simulations, finding that simulations tend to produce MBHs over-massive relative to their host galaxies. Their results indicate simulations typically power too many AGN relative to observations, which may result from generating over-massive MBHs that consistently over-accrete, or may indicate that MBHs in dwarf galaxies are more obscured than previously thought. They suggest that tighter constraints on the AGN fraction from future x-ray facilities may better illuminate the true prevalence of MBHs among low-mass galaxies. However, contamination from x-ray binaries and the hot interstellar medium will likely impact the detection of AGN with total x-ray luminosities fainter than $L_{\rm X}^{\rm AGN} < 10^{38}$ \ergs. Similarly, \citet{Schirra2021} analyze the properties of faint AGN among four large-scale cosmological simulations, finding that the properties of low-luminosity AGN hosts differ strongly between simulations. Their results show that the population of low-luminosity AGN in some simulations are powered by MBHs in massive galaxies $\left(M_{\rm star} > 10^{10} M_\odot\right)$, while in other simulations they are powered by lower-mass MBHs in low-mass galaxies $\left(M_{\rm star} < 10^{10} M_\odot\right)$. These differences may be attributable to differing efficiencies of AGN feedback within each simulation. Regardless, nearly all simulations overestimate the total x-ray luminosity (AGN + non-AGN emission) in star-forming galaxies relative to observations.
    
    In this work, we explore the properties of actively growing MBHs in dwarf galaxies out to $z=2$ in the \Rom{} cosmological simulation. We select dwarf galaxies between $10^{8} M_\odot < M_{\rm star} < 10^{10} M_\odot$, straddling the mass threshold below which the effects of SN feedback are often thought to dominate over AGN feedback, $M_{\rm star} \sim 10^{9} M_\odot$ \citep[e.g,][]{Habouzit2017}. \Rom{} is one of the rare examples of large, high resolution, cosmological hydrodynamic simulations capable of resolving the evolution of dwarf galaxies as small as $M_{\rm star} \gtrsim 10^{7} M_\odot$, while also modelling MBH growth and dynamics within these galaxies. The {\sc TNG-50} \citep{Nelson2019,Pillepich2019} and {\sc FABLE} \citep{Henden2018} simulations reach comparable resolution, but a simplistic MBH seeding mechanism that will only track MBHs in more massive dwarfs at late times. These simulations also do not realistically follow the dynamical evolution of MBHs. The {\sc NewHorizon} simulation \citep{Volonteri2020,Dubois2021}, which re-simulates a $\left(16\;{\rm Mpc}\right)^3$ region of {\sc Horizon-AGN} \citep{Dubois2014} at higher resolution, is most comparable to \Rom{} as it allows MBHs to exist widely within well-resolved, low-mass galaxies while also implementing a prescription for gas dynamical friction (though, see \citet{Pfister2019} for a discussion on the effectiveness of gas dynamical friction relative to that of stars and dark matter). Our analysis provides insight into the prevalence of MBHs and their emitting characteristics across time. With this information, we may further learn more about the impact AGN may have on the evolution of dwarf galaxies.
    
    In Section \ref{Simulation} we describe the physics of the \Rom{} cosmological simulation, including the resultant MBH occupation fraction. In Section \ref{scaling-relations} we consider local scaling relations between $L_{\rm X}^{\rm AGN}$, $M_{\rm star}$ and SFR. In Section \ref{active-fraction} we explore the evolution of active fractions across time. In Section \ref{hidden-mbhs}, we predict a population of MBHs with low x-ray luminosities and high contamination by x-ray binaries and the hot interstellar medium. In Section \ref{number-density} we report on the number densities of MBHs, both hidden and visible, as well as AGN out to $z=2$. We summarize our findings in Section \ref{Conclusion}.

\section{Simulation} \label{Simulation}

    \begin{figure*}
        \plotone{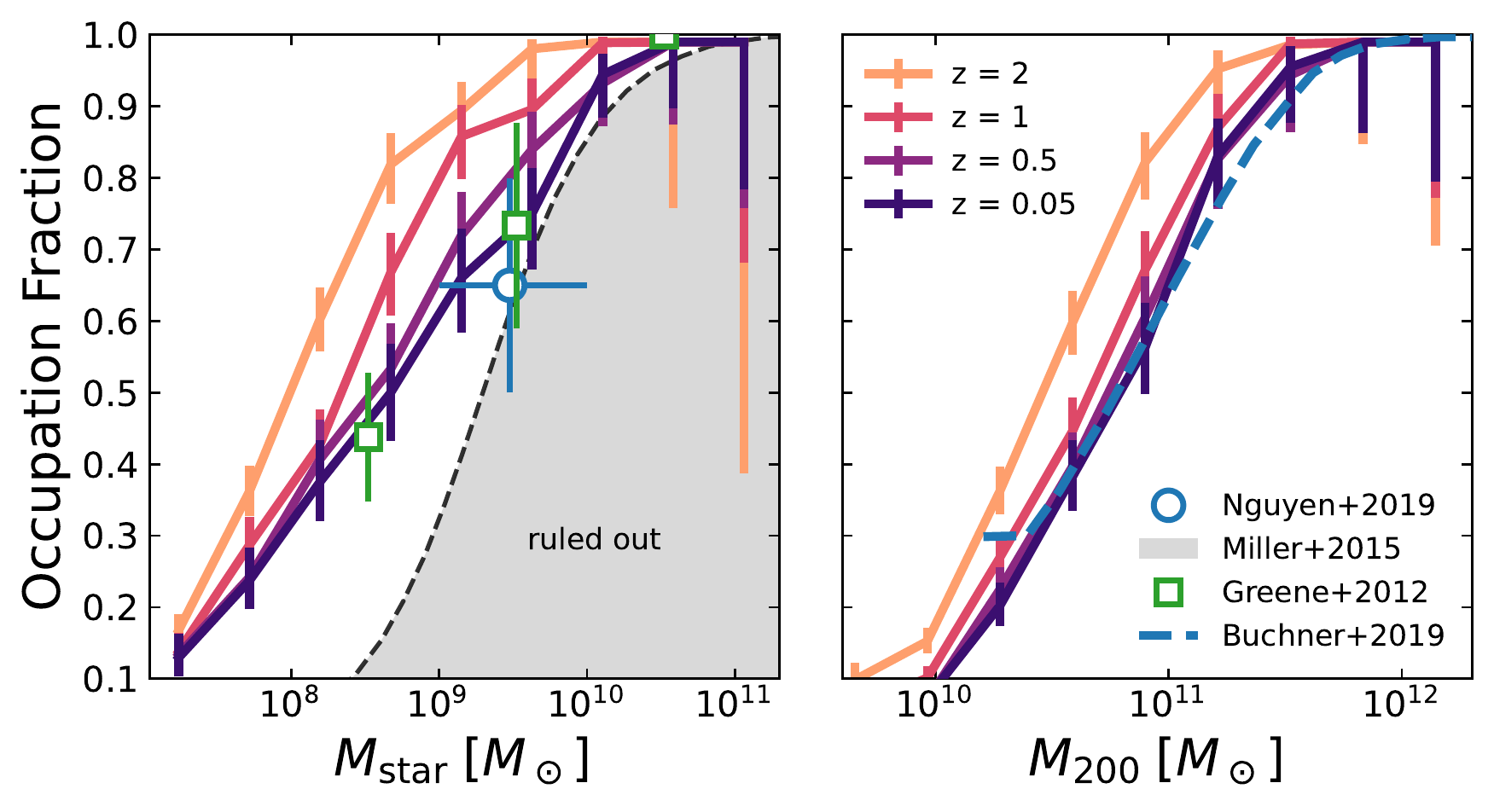}
        \caption{Fraction of galaxies hosting a MBH within $10$ kpc of the galaxy center, as a function of host stellar mass (left) and host halo mass (right), binned by redshift. Error bars indicate $95\%$ binomial uncertainties. We mark observational constraints of the occupation fraction from dynamical MBH estimates \citep[][blue circles]{Nguyen2019} and x-ray observations of optically selected AGN \citep[][green squares]{Desroches2009,Greene2012}. We mark occupation fractions ruled out by x-ray selected AGN \citep[][gray shaded]{Miller2015}. We include constraints of the occupation fraction versus halo mass from analytic models \citep[][blue dashed]{Buchner2019}. At $z=2$, all halos above $M_{\rm star} > 2\times10^{10} M_\odot$, $M_{\rm 200} > 2\times10^{11} M_\odot$ host a MBH within the $10$ kpc of the center. By $z=0.05$, these thresholds shift to $M_{\rm star} > 3\times10^{10} M_\odot$, $M_{\rm 200} > 3\times10^{11} M_\odot$.}
        \label{occupation-fraction}
    \end{figure*}
    
    We now summarize the relevant aspects of the \Rom{} cosmological hydrodynamic simulation. See \citet{Tremmel2017} for a full description of the physical prescriptions.
    
    The \textsc{Romulus} suite of cosmological hydrodynamic simulations, including \Rom{} \citep{Tremmel2017} and \textsc{RomulusC} \citep{Tremmel2019}, were run with the N-Body + Smooth Particle Hydrodynamics (SPH) code \textsc{ChaNGa} \citep{Menon2015, Wadsley2017}. In this work, we focus on the \Rom{} ($25$ Mpc)$^{3}$ uniform-resolution volume. In \Rom{}, dark matter particles have a mass of $3.39 \times 10^{5} M_\odot$, gas particles have a mass of $2.12 \times 10^{5} M_\odot$, and particles have a Plummer equivalent force softening of $250$ pc. \Rom{} contains $3.375\times$ more dark matter particles than gas particles to better resolve MBH dynamics \citep{Tremmel2015}. The simulations were run with a Planck 2014 $\Lambda$CDM cosmology, with $\Omega_m = 0.3086$, $\Omega_\Lambda = 0.6914$, $h = 0.6777$, and $\sigma_8 = 0.8288$, \citep{PlanckCollaboration2014}. 
    
    Throughout our analysis, we adjust simulated stellar masses with corrections from ~\citet{Munshi2013} that account for the impact of aperture photometry on observed stellar masses, such that $M\textsubscript{star, obs} = 0.6 M\textsubscript{star, sim}$. These corrections allow us to perform a more direct comparison between simulated and observed stellar masses. \Rom{} resolves galaxies down to corrected stellar masses $M_{\rm star} \gtrsim 10^{7} M_\odot$.
    
    \subsection{Star Formation}
    
        Stars in \Rom{} form according to a star formation efficiency when cold gas has a density that exceeds a threshold for star formation. 
        \begin{itemize}
            \item Star formation efficiency $c_\ast = 0.15$,
            \item Gas density threshold $n_\ast \ge 0.2$ m$_p$ cm$^{-3}$,
            \item Gas temperature $T < 10^{4}$ K.
        \end{itemize}
        Additionally, $0.75 \times 10^{51}$ erg of thermal energy is deposited in the ISM by SN\,II following the `blastwave' mechanism \citep{Stinson2006}. Cooling is shut off in the gas particles that receive SN\,II energy for a time period representing the adiabatic expansion phase of a blastwave SN remnant.
        
        These star formation parameters were tuned using a set of $80$ zoom-in simulations of $4$ halos with halo masses $M_{\rm vir} = 10^{10.5} - 10^{12} M_\odot$. A set of parameters were chosen by their ability to reproduce $z=0$ scaling relationships between stellar mass, halo mass, \HI gas mass, and angular momentum \citep{Moster2013, Obreschkow2014, Cannon2011, Haynes2011}. 
        
        Prescriptions for metal diffusion \citep{Shen2010}, thermal diffusion \citep{Wadsley2017}, and low-temperature radiative cooling \citep{Guedes2011} are also included in \Rom{}, with a \citet{Kroupa2001} initial mass function for stars. \Rom{} does not include high temperature metal cooling, which should not have an effect in low mass halos with typically low metallicity gas (see \citet{Tremmel2019} for a more detailed discussion).
    
    \subsection{Black Hole Physics}
    
        MBHs in \Rom{} are seeded according to local, pre-collapse gas properties with thresholds and seed mass similar to a direct-collapse model \citep{Haiman2013, Greene2020}. A star-forming gas particle is instead marked to form a MBH if it meets the following criteria:
        \begin{itemize}
            \item Low metallicity, $Z < 3\times10^{-4}$,
            \item Gas density threshold, $n_{\ast, \rm BH} > 3$ m$_p$ cm$^{-3}$,
            \item Temperature, $T = 9500-10^4$ K,
        \end{itemize}
        effectively restricting BH creation to high-density regions in the early universe. If these criteria are met, the gas particle forms an MBH with mass $M_{\rm BH} = 10^{6} M_\odot$, accreting mass from nearby gas particles. This seed mass is somewhat higher than theoretical estimates \citep[e.g,][]{Volonteri2012}. The high seed mass for MBHs allows us to well-resolve their dynamics over cosmic time \citep[][see below]{Tremmel2015}. Additionally, the early growth onto MBH progenitors may exceed $0.1 M_\odot$ yr$^{-1}$, and is governed by processes below the resolution limits of the simulation \citep{Hosokawa2013,Schleicher2013}. In \Rom{}, $95\%$ of MBHs form within the first Gyr of the simulation \citep{Tremmel2017}.
        
        \Rom{} employs prescriptions for MBH dynamical friction that produce realistic MBH sinking timescales \citep{Tremmel2015}. This sub-grid model allows MBHs in large halos to stay centered \citep{Kazantzidis2005,Pfister2017}, but also allows some MBHs to ``wander'' within sufficiently shallow potentials, as has been discovered in recent observations \citep{Reines2020} and simulations \citep[e.g,][]{Tremmel2018,Bellovary2019,Ricarte2021,Ricarte2021a,Bellovary2021}. The dynamical friction prescriptions along with the ``oversampling'' of dark matter particles, the high seed mass, and high resolution of \Rom{} together help avoid unrealistic numerical heating of MBHs and help ensure accurate MBH dynamics.
        
        MBH feedback takes the form of thermal energy injection into the surrounding environment. Thermal energy from the MBH, $E_{\rm BH}$, is isotropically injected into the $32$ nearest gas particles in some time, $dt$, following:
        \begin{align}
            E_{\rm BH} = \epsilon_r \epsilon_f \dot{M} c^2 dt,
        \end{align}
        where $\epsilon_r = 0.1$ is the radiative efficiency, $\epsilon_f = 0.02$ is the energy injection efficiency, and $\dot{M}$ is the MBH accretion rate. 
        
        Accretion itself follows a modified Bondi-Hoyle prescription to incorporate angular momentum on unresolved spatial scales. The ``instantaneous'' accretion is averaged over the smallest simulation time element, typically $10^{4} - 10^{5}$ yr, and remains Eddington limited at all times. We can write the accretion rate depending on whether the dominant gas motion is rotational, $v_{\theta}$, or bulk flow, $v\textsubscript{bulk}$:
		\begin{align}
			\dot{M}= \alpha \times
				\begin{cases}
					\frac{\pi G^2 M^2_{\rm BH} \rho}{\left(v_{bulk}^2 + c_s^2\right)^{3/2}} &\text{if } v_{bulk} > v_\theta\\
					\frac{\pi G^2 M^2_{\rm BH} \rho c_s}{\left(v_\theta^2 + c_s^2\right)^2} &\text{if } v_{bulk} < v_\theta,
				\end{cases}
		\end{align}
		where
		\[
			\alpha =
				\begin{cases}
				   \left(\frac{n\textsubscript{gas}}{n\textsubscript{*}}\right)^\beta &\text{if } n\textsubscript{gas} \geq n\textsubscript{*}\\
				   1 &\text{if } n\textsubscript{gas} \leq n\textsubscript{*},
				\end{cases}
		\]
		is the density-dependent boost factor that corrects for underestimated accretion rates due to resolution limitations \citep{Booth2009}, $\beta = 2$ is the corresponding boost coefficient, $n\textsubscript{gas}$ is the number density of the surrounding gas, $n\textsubscript{*}$ is the star formation density threshold, $\rho$ is the mass density of the surrounding gas, $c\textsubscript{s}$ is the local sound speed, $v_\theta$ is the rotational velocity of the surrounding gas at the smallest resolved scales, and $v\textsubscript{bulk}$ is the bulk velocity of the surrounding gas. This calculation is performed over the 32 nearest particles. Free parameters related to accretion and feedback were optimized to reproduce various empirical scaling relations for low mass halos $\left(M_{\rm vir} \lesssim 10^{12} M_\odot\right)$ at $z=0$ \citep{Tremmel2017}, including the observed relationship between MBH mass and stellar mass \citep{Schramm2013}.
        
        Figure \ref{occupation-fraction} illustrates the evolution of the MBH occupation fraction in \Rom{}, where the occupation fraction is defined as the fraction of halos of a given stellar/halo mass containing at least one MBH within the inner $10$ kpc. We do not include MBHs within substructure of the primary halo. The high-redshift MBH occupation fraction is determined by the seeding mechanism, while the time evolution is primarily driven by structure growth. We show the occupation fraction as both a function of stellar mass and halo mass. At $z=0.05$, the MBH occupation fraction drops below unity at $M_{\rm star} = 3\times10^{10} M_{\odot}$, $M_{\rm 200} = 3\times10^{11} M_\odot$.
        
        The local $z=0.05$ occupation fraction is consistent with constraints from observations and empirical modelling. Dynamical MBH mass estimates \citep{Nguyen2019} place the occupation fraction between $50-80$\%  for dwarf galaxies between $10^{9} M_\odot < M_{\rm star} < 10^{10} M_\odot$. X-ray selected AGN \citep{Miller2015} provide lower limits on the occupation fraction as a function of stellar mass. \citet{Greene2012} estimate the occupation fraction using x-ray observations of $8$ optically selected AGN with $M_{\rm BH} > 3\times10^5 M_\odot$, as estimated from the $M_{\rm BH} - \sigma$ relation \citep{Desroches2009}. \citet[][not shown]{Baldassare2020} use variability of optical light curves from the NASA–Sloan Atlas with Palomar Transient Factory coverage to identify local, low-mass AGN. They find that AGN variability fractions are approximately constant down to stellar masses $M_{\rm star} = 10^{9} M_\odot$, suggesting the MBH occupation fraction does not change much between $M_{\rm star} > 10^{9-10} M_\odot$, consistent with our estimates. The local occupation fraction also agrees with occupation fractions from \citet[$p=0.3, \log M_c = 10$]{Buchner2019}, who use empirical models to explore valid regions of the parameter space of critical halo mass versus MBH seed probability.
        
        \subsection{Halo and Galaxy Extraction}
        
        We use the Amiga Halo Finder \citep{Knollmann2009} to extract gravitationally bound dark matter halos, as well as sub-halos, and the baryonic content associated with these structures. AHF utilizes a spherical top-hat collapse technique to define the virial radius (R$_{\mathrm{vir}}$) and mass (M$_{\mathrm{vir}}$) of each halo and sub-halo. AHF uses a spherical top-hat collapse technique to define the virial radius and total mass of each halo. The center of each halo/galaxy is calculated using a shrinking spheres approach \citep{Power2003}.
    
\section{Results} \label{Results}

    In this work, we are primarily interested in dwarf galaxies between corrected stellar masses $10^{8} M_\odot < M_{\rm star} < 10^{10} M_\odot$ at redshifts $z=0.05 - 2$. Aside from encompassing the regime where feedback is thought to change from AGN-dominated to SN-dominated \citep{Habouzit2017}, these limits are also similar to the ranges for dwarf AGN observed to date \citep[e.g,][]{Mezcua2018,Birchall2020,Birchall2022}.
    
    Galaxies in \Rom{} frequently host multiple MBHs, and some galaxies have been found to host ``wandering'' MBHs at large radii from the center \citep{Tremmel2018,Ricarte2021a,Ricarte2021}. We restrict our sample to MBHs within $10$ kpc of the halo center, similar to the optical counterpart search radius from \citet{Birchall2020}. In cases where there are multiple MBHs within the central region, we choose the most {\it luminous}. This choice is nearly always ($> 95$\% of the time) the same as choosing the most massive or most central MBH.
	
	At low stellar masses in \Rom{}, MBH seed masses often make up the majority of the total MBH mass. Further, the simulation does not impose restrictions on seeding multiple MBHs within close proximity, sometimes allowing multiple seeds to rapidly merge at high redshift. \citep{Ricarte2019}. When reporting on MBH masses, we subtract off the seed masses of all progenitor MBHs from the final mass such that:
	\begin{equation}
	    M_{\rm BH}^{\rm acc} = M_{\rm BH} - M_{\rm seeds}.
	\end{equation}
	Subtracting off all progenitor seed masses maintains the accreted mass from MBHs merging onto the main progenitor, but removes the contribution from seeding. \citet{Ricarte2019} find that MBHs in \Rom{} with total masses $>10^{7} M_\odot$ are dominated by accretion, and that the accreted MBH mass follows the observed $M_{\rm BH} - M_{\rm star}$ relation down to dwarf galaxy scales. This result suggests that the accreted MBH mass may be a more realistic proxy for the true MBH mass within dwarf galaxies in \Rom{}, \citep[although, see][for a discussion on MBH accretion properties in high resolution dwarf galaxy simulations.]{Bellovary2019}. It is important to note that without these adjustments, the total MBH masses in low-mass \Rom{} galaxies are over-massive compared to observed scaling relations \citep{Sharma2020}, which indicates the seed masses are unrealistically large. It is possible that the fiducial Bondi accretion model within \Rom{} is unsuitable for modeling accretion in dwarf galaxies -- an alternate accretion model that allows for the same growth with lower seed masses may be more realistic.
	
	In order to calculate bolometric luminosities for MBHs, we follow the \citet{Churazov2005} two-mode accretion model that distinguishes between radiatively efficient and inefficient AGN. Many simulations, including \Rom{}, use a single, radiatively efficient accretion model for internally calculating feedback \citep{Hirschmann2014}. Using a two-mode accretion model in post-processing reduces the number of low-luminosity AGN, though it misses the feedback effects of radiatively inefficient AGN. We convert the instantaneous accretion rate, $\dot{M}_{\rm BH}$, such that:
	\begin{equation}
	    L_{\rm bol} = \begin{cases}
    	    \epsilon_r \dot{M}_{\rm BH} c^2, \quad & f_{\rm Edd} \geq 0.1\\
    	    10 f_{\rm Edd} \epsilon_r \dot{M}_{\rm BH} c^2, \quad & f_{\rm Edd} < 0.1.
	    \end{cases}
	\end{equation}
	with radiative efficiency $\epsilon_r = 0.1$ and Eddington fraction $f_{\rm Edd} = \dot{M}_{\rm BH}^{\rm acc}\,/\,\dot{M}_{\rm Edd}$. Note that we calculate the Eddington fraction using the accreted MBH mass, $M_{\rm BH}^{\rm acc}$, and hence $f_{\rm Edd}$ is {\it higher} than when calculated using the total MBH mass. In the two-mode accretion model, higher $f_{\rm Edd}$ leads to higher luminosities (and hence higher, more conservative estimates of the active fractions in Section \ref{active-fraction}) among radiatively inefficient AGN. To estimate x-ray luminosities, $L_{\rm X}^{\rm AGN}$, between $0.5 - 10$ keV, we apply \citet{Shen2020} bolometric corrections in the soft $\left(0.5 - 2\;{\rm keV}\right)$ and hard $\left(2 - 10\;{\rm keV}\right)$ x-ray bands.
        
    As done in \citet{Koudmani2021,Haidar2022}, we calculate mock x-ray luminosities for dwarf galaxies by modeling the contributions from x-ray binaries (XRBs) and emission from the hot interstellar medium. We model high- and low-mass x-ray binary contributions using the \citet{Lehmer2016} relation:
    \begin{align}
        L_{\rm X}^{\rm XRB} = \alpha_0 \left(1 + z \right)^\gamma M_{\rm star} + \beta_0 \left(1 + z\right)^\delta {\rm SFR},
    \end{align}
    where $\left(\log\alpha_0, \gamma, \log\beta_0, \delta\right) = \left(29.04, 3.78, 39.38, 0.99\right)$ for soft x-rays, and $\left(29.37, 2.03, 39.28, 1.31\right)$ for hard x-rays. Emission from hot gas follows the \citet{Mineo2012} relation, which establishes a relationship between ($0.5 - 2$ keV) x-ray luminosity of the diffuse interstellar medium and the star formation rate:
    \begin{align}
        L_{\rm X, soft}^{\rm gas} = 8.3 \times 10^{38} \times \left(\frac{\rm SFR}{M_\odot\,{\rm yr}^{-1}}\right).
    \end{align}
    We calculate the hot gas contribution in the hard x-ray band by assuming a photon index $\Gamma = 3$. Hence we can write the total ($0.5-10$ keV) x-ray luminosity for a galaxy:
    \begin{align}
        L_{\rm X} = L_{\rm X}^{\rm AGN} + L_{\rm X}^{\rm XRB} + L_{\rm X}^{\rm Gas}.
    \end{align}
    
    In order to more closely compare with observations from \citet{Mezcua2018} and \citet{Birchall2020}, we classify AGN in \Rom{} such that their x-ray luminosity is significantly greater than the contribution from non-AGN sources, $L_{\rm X}^{\rm AGN} \geq 2 L_{\rm X}^{\rm XRB+Gas}$. We do not model emission from background quasars, which have been found to contaminate observations \citep[e.g,][]{Reines2020}.
    
    Recent work by \citet{Kristensen2021} indicates that among low-mass $\left(10^{9} < M_{\rm star} < 3\times10^{9} M_\odot\right)$ galaxies in three large-scale cosmological simulations, non-AGN are more likely than AGN to be found in a denser environment with closer galactic neighbors. Their results also indicate that galaxies which maintain close proximity to other galaxies are less likely to exhibit star formation or AGN activity. While we do not show results from subhalos of the target halo in this work, we do not make further selections based on galaxy environment. Although we lose the ability to identify the impact of galaxy environment, we are able to make more direct comparisons with \citet{Mezcua2018} and \citet{Birchall2020}, who similarly do not make selections on galaxy environment.
        
	\subsection{Dwarfs in the local universe} \label{scaling-relations}
        
        \begin{figure*}
            \plotone{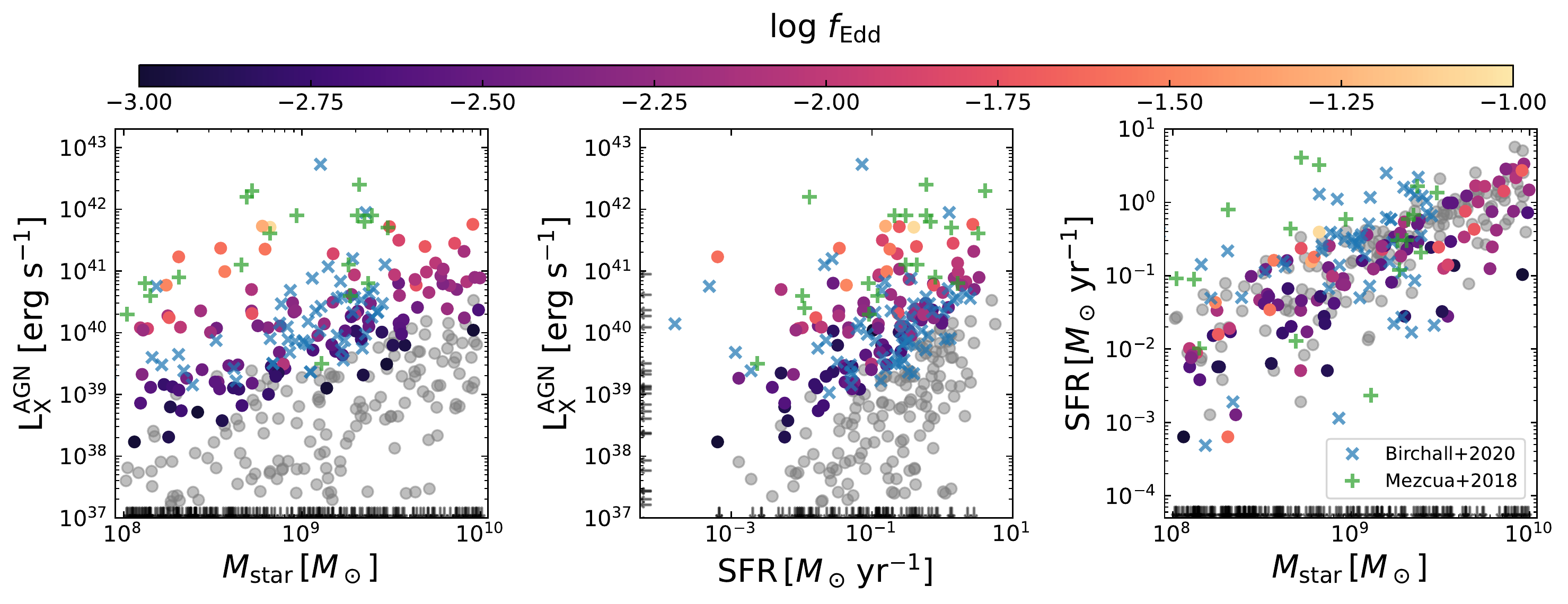}
            \caption{Local $z=0.05$ scaling relations colored by instantaneous Eddington fraction. Galaxies that fall below the contamination threshold, $L_{\rm X}^{\rm AGN} < 2 L_{\rm X}^{\rm XRB+Gas}$, are marked in gray. We include x-ray detected AGN from \citet[][blue $\times$]{Birchall2020} and \citet[][green $+$]{Mezcua2018} between $10^{8} < M_{\rm star} < 10^{9.5} M_\odot$. Uncontaminated \Rom{} dwarfs are in agreement with the observed $L_{\rm X}^{\rm AGN} - M_{\rm star}$ relation (left), $L_{\rm X}^{\rm AGN} - {\rm SFR}$ relation (middle), and ${\rm SFR} - M_{\rm star}$ relation for local dwarf AGN. Contaminated galaxies do not follow observed $L_{\rm X}^{\rm AGN} - M_{\rm star}$ or $L_{\rm X}^{\rm AGN} - {\rm SFR}$ relations, suggesting a breakdown for undetected MBHs in dwarf galaxies.}
            \label{observability-plot}
        \end{figure*}
    
        Dwarf galaxies have been found to exhibit relationships between MBH x-ray luminosity, stellar mass, and star formation rate (SFR) similar to those found in massive galaxies \citep[e.g,][]{Aird2017,Aird2018}. These relationships may illuminate any connection between MBH activity and star formation, as well as provide insight into how readily dwarf AGN at each mass scale may be observed.
        
        Figure \ref{observability-plot} shows the relationships between $L_{\rm X}^{\rm AGN}$, $M_{\rm star}$, and SFR for dwarf galaxies at $z=0.05$ that host an MBH in the central $10$ kpc. Each point that satisfies the contamination threshold $L_{\rm X}^{\rm AGN} > 2L_{\rm X}^{\rm XRB+Gas}$ is colored according to the instantaneous Eddington fraction. Dwarfs that fail the contamination threshold are marked in gray. Arrows mark dwarfs with extremely low luminosities or zero star formation, with star formation rates averaged over the past $100$ Myr. 
        
        We compare our results with x-ray selected dwarf AGN from \citet{Mezcua2018}, who identify $40$ dwarf AGN between $10^{7} M_\odot < M_{\rm star} < 3 \times 10^{9} M_\odot$ from the {\it Chandra COSMOS-Legacy} survey out to $z<2.4$, in the full $0.5 - 10$ keV band. We only show their sample up to $z < 0.25$ for better comparison with our local dwarfs. We also include results from \citet{Birchall2020} who provide $61$ x-ray selected dwarf AGN that fall within both the optical MPA-JHU footprint and x-ray 3XMM footprint. \citet{Birchall2020} identify dwarfs with $M_{\rm star} < 3 \times 10^{9} M_\odot$ out to $z < 0.25$ in the harder $2 - 12$ keV band. Both observational sets calculate the AGN x-ray luminosity by subtracting off contributions to the total x-ray luminosity from XRBs and the hot interstellar medium.
        
        Overall, the scaling relationships between our uncontaminated luminous dwarfs are consistent with observed relations of local dwarf AGN. Simulated dwarfs tend not to reach the highest luminosities found in the observations, $L_{\rm X}^{\rm AGN} > 10^{42}$ \ergs, though these observed luminous dwarfs are typically at slightly higher redshifts, $z\gtrsim0.1$. Our dwarfs strike a middle ground between the higher luminosities found in \citet{Mezcua2018} and the lower luminosities found in \citet{Birchall2020}, which are at least partially due to the harder x-ray band in which \citet{Birchall2020} observe.
        
        There exists a large population of dwarfs at low luminosities and high contamination ($L_{\rm X}^{\rm AGN} < 2L_{\rm X}^{\rm XRB+Gas}$) that are not found in observations. These contaminated dwarfs make up $70\%$ of dwarfs with MBHs at $z=0.05$. This hidden population that is missed by observations suggests that typical relationships between $L_{\rm X}^{\rm AGN}$ and both $M_{\rm star}$ and SFR are dramatically impacted by a survey's ability to detect low luminosities and distinguish contaminated AGN. Indeed the largest scatter in these relationships are found at dwarf galaxy masses, suggesting the relationships found in massive galaxies may break down for dwarf galaxies. We further explore the role of detection threshold and x-ray contaminants in Section \ref{hidden-mbhs}.
        
        Many simulated dwarfs also exhibit quenched star formation. Since MBH accretion is, on average, correlated with SFR among star-forming main sequence galaxies in \Rom{} \citep{Ricarte2019}, it is unsurprising that many low-luminosity MBHs are found in galaxies with little star formation. However, a surprising number of quenched dwarfs exhibit an actively accreting MBH, in-line with x-ray observations from \citet{Aird2019,Carraro2020}, which both find elevated AGN activity among quiescent galaxies relative to star-forming galaxies at the same SFR. This phenomenon suggests the mechanism that fuels star formation is, at least in some dwarfs, separate from that which fuels MBH activity. It may also indicate a connection between active accretion and the suppression of star formation in some dwarfs.
        
        It is worth noting that $20\%$ of quenched (zero SFR) dwarf galaxies in \Rom{} are isolated, meaning they are (i) outside the virial radius of a larger halo, and (ii) farther than $>1.5$ Mpc from any neighboring galaxy with $M_{\rm star} > 2.5 \times 10^{10} M_\odot$ \citep{Geha2012}. The high number of quenched, isolated dwarf galaxies in \Rom{} is in tension with observations of local field dwarfs \citep{Geha2012} where the quenched fraction in isolated dwarf galaxies is $f_{\rm q} \sim 0.01 - 0.1$.  \citet{Dickey2021} find that many other cosmological simulations similarly over-produce quenched, isolated dwarfs. Future work will explore the dwarf quenching mechanism in \Rom{}, including the possible ties to AGN activity.
        
    \subsection{Active fraction of detectable MBHs} \label{active-fraction}
        
           \begin{figure}
            \plotone{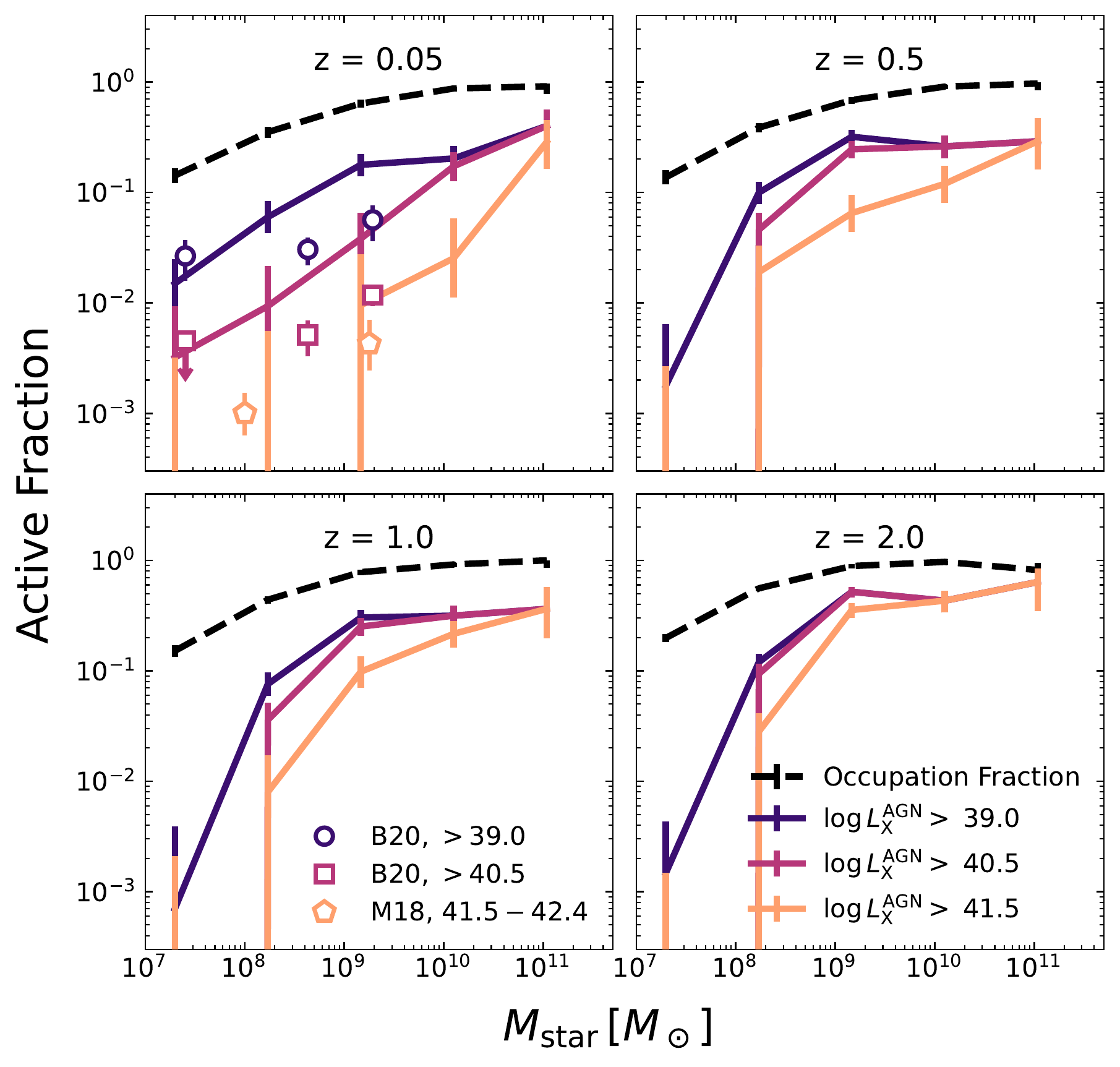}
            \caption{Fraction of all galaxies containing an active MBH versus stellar mass, in bins of redshift. Activity is defined by three x-ray luminosity thresholds, $\log L_{\rm X}^{\rm AGN} > [39, 40.5, 41.5]$}. The MBH occupation fraction (black dashed) sets upper limits on the active fraction. Error bars mark $95\%$ binomial uncertainties. Active fractions peak at $z=2$ and decrease with time, at all stellar masses below $M_{\rm star} < 10^{11} M_\odot$. Massive galaxies exhibit consistently higher active fractions than low-mass galaxies at all times. At $z=0.05$, $4\%$ of dwarfs emit above $L_{\rm X}^{\rm AGN} > 10^{40.5}$ \ergs, a factor of a few higher than found in x-ray observations from \citet[][circles, squares]{Birchall2020} and \citep[][ pentagons]{Mezcua2018}.
            \label{active-fraction-mstar}
        \end{figure}
        
        Next, we examine the evolution of AGN prevalence across cosmic time. The active fraction is defined here as the fraction of all galaxies emitting above a given $0.5-10$ keV x-ray luminosity threshold. Figure \ref{active-fraction-mstar} shows the active fraction for all galaxies in \Rom{} above $M_{\rm star} > 10^{8} M_\odot$ as a function of stellar mass, in bins of redshift. We define activity using three x-ray luminosity thresholds and include observational constraints from x-ray observations \citep{Mezcua2018,Birchall2020}. We set a cut on acceptable contamination such that $L_{\rm X}^{\rm AGN} > 2 L_{\rm X}^{\rm XRB+Gas}$. Setting this cut on contamination impacts the active fractions in lower luminosity AGN, but does not change fractions among the two highest AGN luminosities shown here.
        
        In addition to expressing a lower occupation fraction than massive galaxies, low-mass galaxies are overall less luminous and hence exhibit a lower active fraction. While the most massive galaxies consistently emit well above $L_{\rm X}^{\rm AGN} > 10^{41.5}$ \ergs, only the higher mass dwarf galaxies between $10^{9} < M_{\rm star} < 10^{10} M_\odot$ reach such high luminosities, and most prominently around $z=2$. Dwarf galaxies peak in activity around $z=2$ and drop off in luminosities with time, steeply dropping around $z=0.05$. By $z=0.05$, approximately $2\%$ of galaxies between $10^{8} < M_{\rm star} < 10^{9} M_\odot$, and $8\%$ of galaxies between $10^{9} < M_{\rm star} < 10^{10} M_\odot$, host an x-ray detectable MBH brighter than $L_{\rm X}^{\rm AGN} > 10^{40.5}$ \ergs. For reference, this luminosity threshold is similar to the detection threshold for the $4.6$ Ms \textit{Chandra-COSMOS-Legacy} observations at $z=0.05$ in the full x-ray band \citep{Suh2017}. Despite the steep drops in luminosity, dwarf galaxies at $z=0.05$ in \Rom{} exhibit a factor of a few higher active fractions than found in x-ray observations.
        
        There may be a few reasons why our dwarf active fractions are higher than the observed. 1) The high seed mass in \Rom{} may ultimately drive unrealistically high accretion rates onto dwarf AGN at the wrong times. Although our dwarf AGN follow observed local scaling relations, including the $M_{\rm BH} - M_{\rm star}$ relation for more massive galaxies, dwarfs in reality may follow a different $M_{\rm BH} - M_{\rm star}$ relation from massive galaxies \citep{Reines2015}. Further, it is possible to correctly predict the empirical relationships while still over-predicting the AGN fraction if the timing of the simulated MBH accretion history is wrong. Indeed, \citet{Ricarte2019} find that the luminosity density of AGN in \Rom{} is high relative to observations at $z=0$. 2) MBHs in \Rom{} accrete on timescales greater than $\gtrsim10^{4}$ yrs. A duty cycle of order $10^{3}$ yrs may account for factor $\sim 10$ differences in observed fraction. 3) Although our occupation fractions are consistent with observations, changing the MBH seeding model to be more restrictive may alleviate the AGN fraction discrepancy while maintaining agreement with observational constraints on the occupation fraction. This change is largely equivalent to changing the seeding model of the simulation. 4) As with other cosmological simulations, we do not directly measure AGN luminosities but instead assume a radiative efficiency to convert MBH accretion rates. We convert accretion rates using a two-mode accretion model where radiative efficiency depends on Eddington fraction \citep{Churazov2005}. However, there is no consensus on the Eddington fraction distributions that underlie accretion models for low luminosity AGN, and hence the precise dependence on Eddington fraction is unclear \citep{Trump2011,Weigel2017,Pesce2021}. Further, we do not take into account AGN obscuration. Although the obscuration fraction among AGN is still uncertain, population synthesis models from \citet{Ananna2019} indicate that $50\%$ of AGN within $z\sim0.1$ may be Compton-thick. 5) Similarly, we rely on existing bolometric corrections from \citet{Shen2020} to calculate x-ray luminosities, but these corrections are calibrated for more massive galaxies. As with radiative efficiency, there is evidence that x-ray bolometric corrections depend on Eddington fraction \citep{Lusso2012}. Radiative efficiency and bolometric corrections are closely linked quantities, and together form a free parameter that is not well constrained within dwarf galaxies \citep{Baldassare2017}. Indeed, \citet{Latimer2021} find evidence that IR-selected dwarf AGN are comparatively under-luminous in the x-ray regime, suggesting a breakdown in typical luminosity scaling relations at dwarf masses, as we have found. Moreover, \citet{Molina2021a} identify $81$ AGN candidates using coronal line emission in the optical regime and find that $49\%$ cannot be correctly identified using other AGN-detection techniques, including x-ray detection.
      
      \begin{figure}
            \plotone{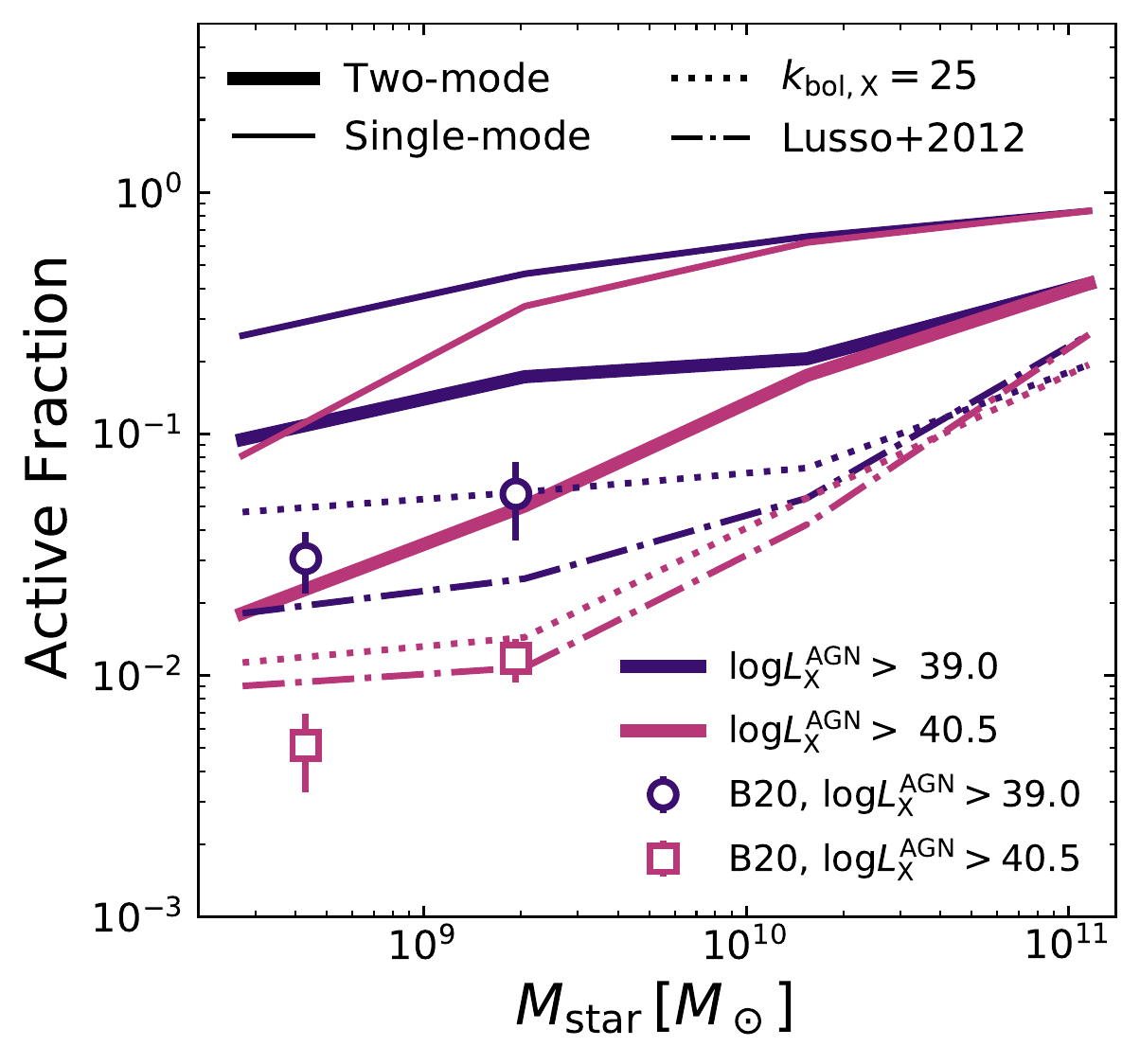}
            \caption{Active fractions versus stellar mass for the typical two-mode accretion model (solid thick), for a single-mode accretion model with a flat radiative efficiency $\epsilon_r = 0.1$ (solid thin), for two-mode accretion with a flat x-ray bolometric correction, $k_{\rm bol, X} = 25$ (dotted), and for two-mode accretion with a luminosity-dependent bolometric correction derived from x-ray selected AGN \citep[][dashed]{Lusso2012}. We show two luminosity thresholds $L_{\rm X}^{\rm AGN} > 10^{39}$ \ergs (purple) and $L_{\rm X}^{\rm AGN} > 10^{40.5}$ \ergs (magenta). A single-mode model yields active fractions $\sim 1$ dex higher than observations \citep[][blue points]{Birchall2020}, while two-mode accretion with a factor of $\sim5$ higher bolometric corrections mitigates the differences above $M_{\rm star} > 10^{9} M_\odot$.}
            \label{active-fraction-diff}
        \end{figure}
        
        Figure \ref{active-fraction-diff} illustrates how the active fraction differs when switching from a two-mode accretion model to a single-mode model in which $L_{\rm bol} = \epsilon_r \dot{M}_{\rm BH} c^2$. We calculate the active fraction for the two-mode model using a flat bolometric correction, $k_{\rm bol, X} = L_{\rm bol}\,/L_{\rm X}^{\rm AGN} = 25$. We also explore the luminosity-dependent bolometric corrections obtained from x-ray selected AGN by \citet{Lusso2012}, applying them to the two-mode accretion model. For the low luminosities found among \Rom{} dwarfs, the extrapolated luminosity-dependent corrections are approximately $k_{\rm bol, X} \sim 10$. \citet{Lusso2012} provide Eddington-fraction bolometric corrections for hard x-rays, but we do not use those corrections since their AGN do not extend down to the Eddington fractions found in \Rom{}.
        
        Single-mode accretion, wherein even MBHs with low Eddington fractions radiate equally efficiently, produces luminosities that are significantly higher than in two-mode accretion. Active fractions for single-mode accretion at a given stellar mass are approximately $1$ dex higher than observations from \citet{Birchall2020}. On the other hand, applying a flat bolometric correction of $k_{\rm bol, X} = 25$ ($\sim 5\times$ larger than corrections from \citealt{Shen2020}) to the two-mode model yields even closer consensus with observations. The luminosity-dependent corrections derived from x-ray AGN yield even lower active fractions, despite having $k_{\rm bol, X} \sim 10$ (only $\sim 2\times$ larger than corrections from \citealt{Shen2020}). It is worth noting that among massive galaxies, high bolometric corrections of order $k_{\rm bol} \sim 100$ are typically only found in Compton-thick AGN and/or AGN with particularly high Eddington fraction \citep{Vasudevan2007,Lusso2012,Brightman2017}, while the dwarfs in \Rom{} typically exhibit Eddington fractions $f_{\rm Edd} < 10^{-2}$.
    
    \subsection{Population of hidden MBHs} \label{hidden-mbhs}
    
        \begin{figure}
            \plotone{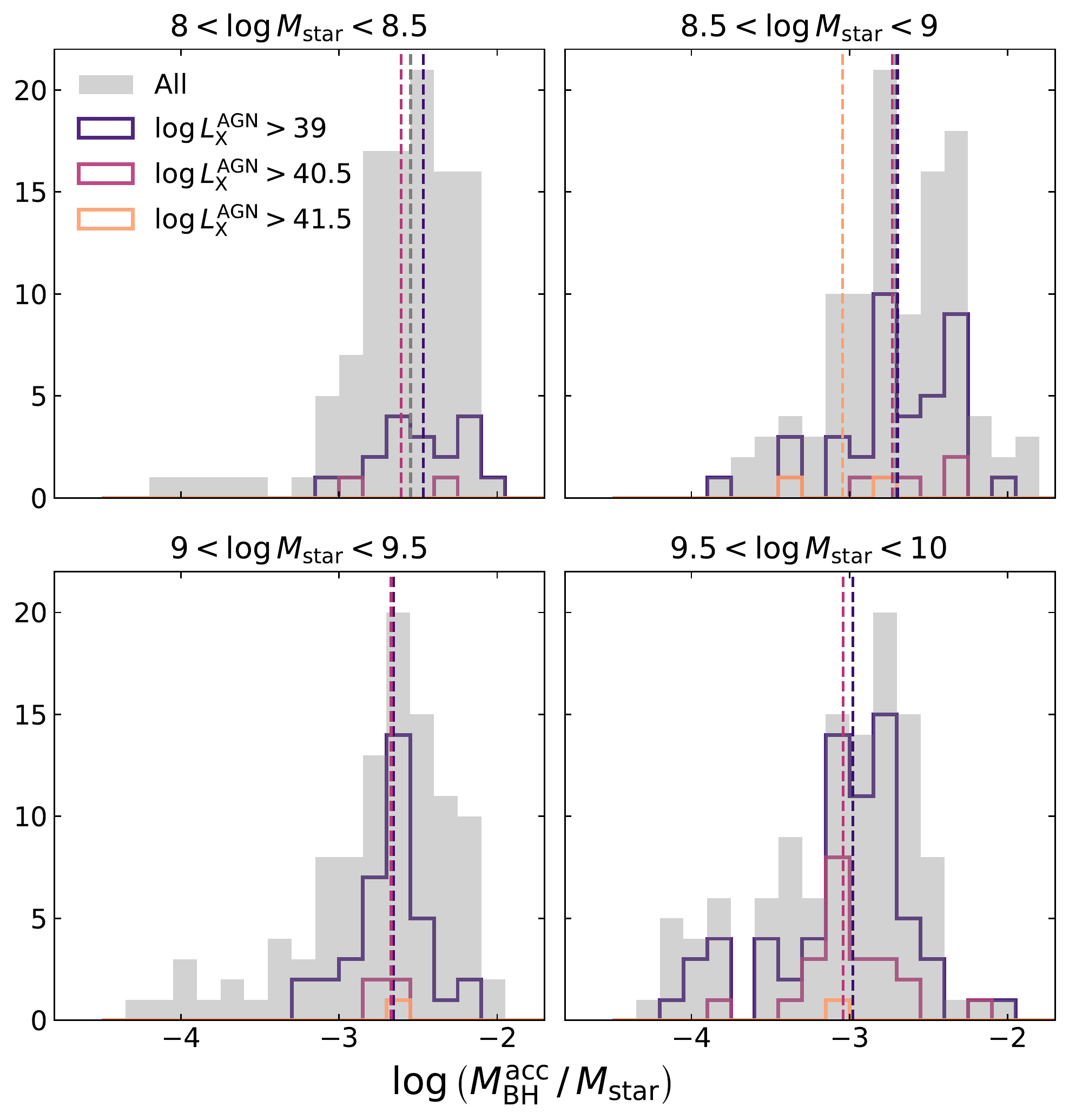}
            \caption{Distribution of MBH-host mass ratios at $z=0.05$ among all MBHs (gray), and for MBHs with x-ray luminosities $L_{\rm X}^{\rm AGN} > 10^{39}$ \ergs (purple), $L_{\rm X}^{\rm AGN} > 10^{40.5}$ (magenta), and $L_{\rm X}^{\rm AGN} > 10^{41.5}$ \ergs (gold), binned by stellar mass. Dashed lines indicate the median of each distribution. Increasing the x-ray detection limits removes most under-massive MBHs from the detected sample in all stellar mass bins. Despite missing under-massive MBHs, raising the detection threshold does not strongly change the underlying distribution of mass ratios.}
            \label{mbh-mstar-hist-luminosity}
        \end{figure}
        
        \begin{figure}
            \plotone{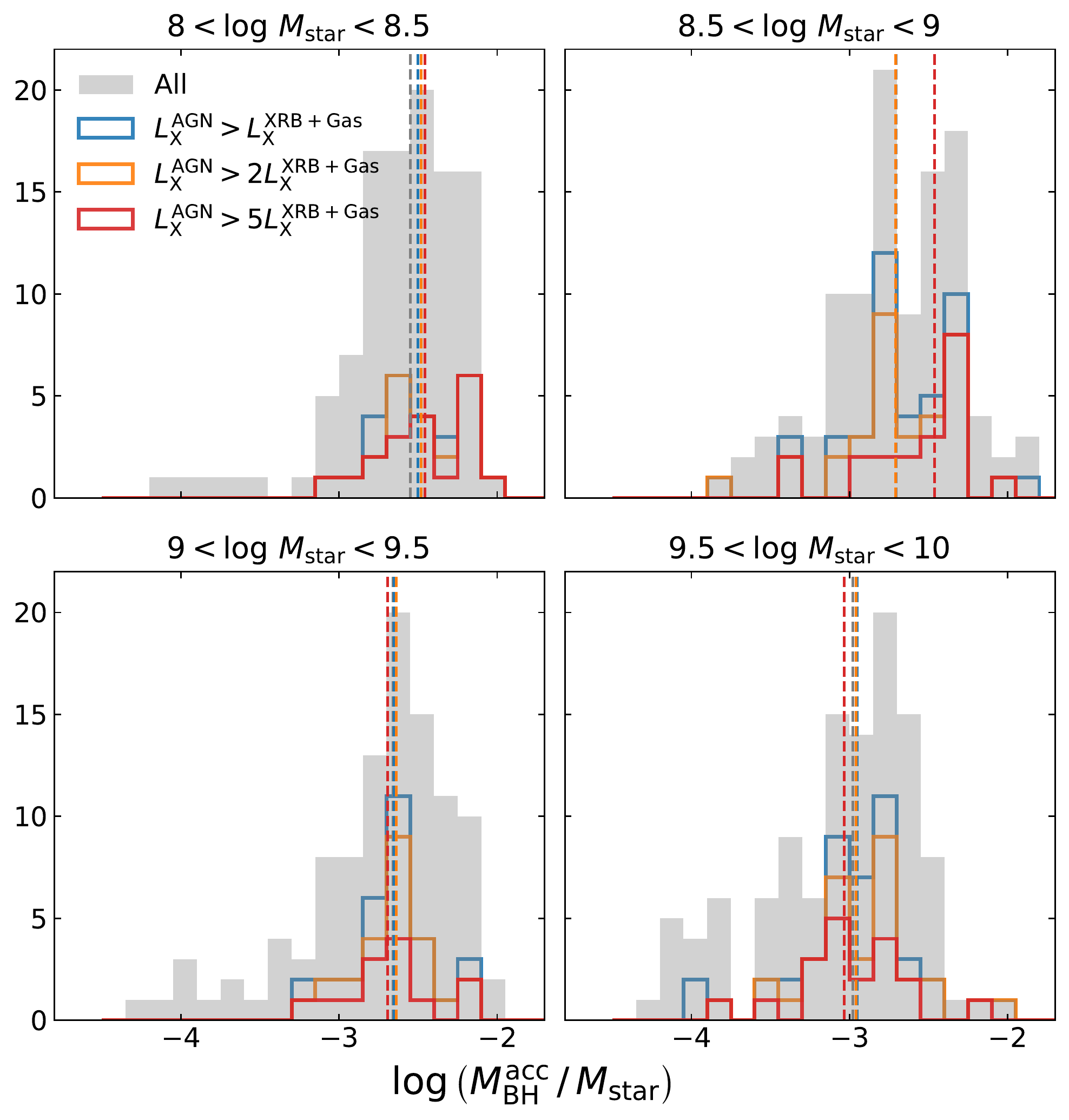}
            \caption{Distribution of MBH-host mass ratios at $z=0.05$ among all MBHs (gray), and for MBHs selected with contamination thresholds $L_{\rm X}^{\rm AGN} > f_{\rm cont}\times L_{\rm X}^{\rm XRB+Gas}$ with $f_{\rm cont} = 1$ (red), $f_{\rm cont} = 2$ (orange), and $f_{\rm cont} = 5$ (blue). Dashed lines indicate the median of each distribution. Selecting AGN with fractions as low as $f_{\rm cont} = 1$ misses most under-massive MBHs. However, raising or lowering the contamination threshold does not strongly change the underlying distribution of mass ratios.}
            \label{mbh-mstar-hist-contamination}
        \end{figure}
        
        \begin{figure*}
            \plotone{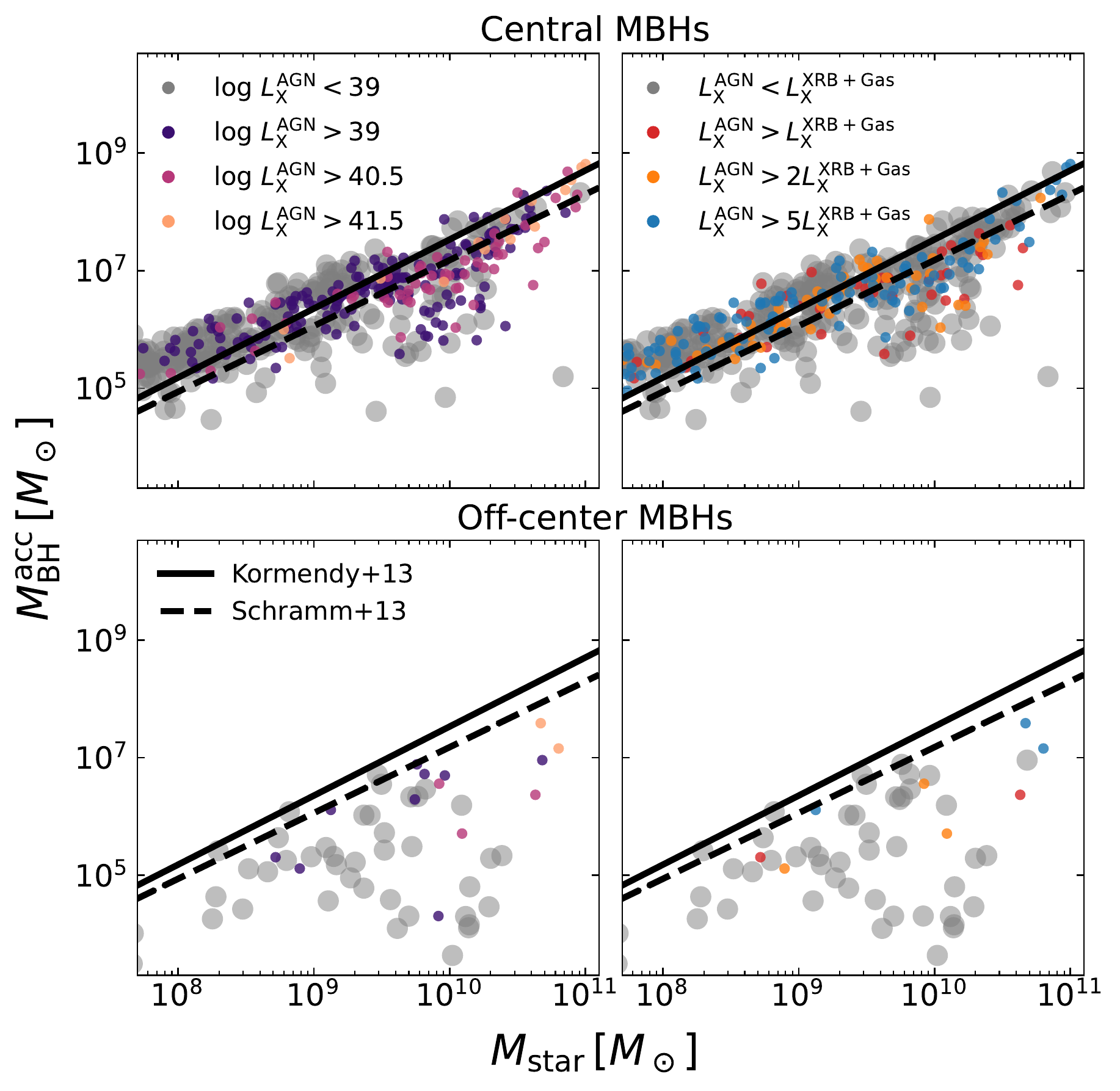}
            \caption{The $M_{\rm BH}^{\rm acc} - M_{\rm star}$ relation for galaxies up to $M_{\rm star} < 10^{11} M_\odot$ at $z=0.05$. We distinguish between the brightest MBHs within $2$ kpc (top) and the brightest MBHs outside of $2$ kpc (bottom). {\it Left}: Relation with cuts on x-ray luminosity and no cuts on contamination fraction. Colors are as in Figure \ref{mbh-mstar-hist-luminosity}. {\it Right}: Relation with cuts on contamination fraction and no cuts on AGN luminosity. Colors are as in Figure \ref{mbh-mstar-hist-contamination}. Central MBHs follow observed relations from \citet[][dashed]{Schramm2013} and \citet[][solid]{Kormendy2013}, while off-center MBHs exhibit higher scatter and encompass many more under-massive MBHs. Setting common detection thresholds and common contamination fractions misses many MBHs in low-mass galaxies as well as the most undermassive MBHs at all stellar masses. Nearly all off-center MBHs are undermassive, low-luminosity, and highly contaminated.}
            \label{mbh-mstar-scatter}
        \end{figure*}
        
        Next, we illuminate a population of MBHs within \Rom{} that may be hidden from current x-ray surveys. We explore the impact of current x-ray detection limits; contamination of low luminosity AGN by XRBs and the hot interstellar medium; and off-center (outside 2 kpc of the galaxy center) versus central (within 2 kpc of the galaxy center) MBHs within dwarf galaxies on the detected population of dwarf AGN.
        
        Using semi-analytic models of MBH growth in low-mass galaxies, \citet{Pacucci2018} find that the tight scaling relationship between MBH mass and bulge stellar velocity dispersion, $\sigma_\star$, found in high-mass galaxies tends to over-predict $M_{\rm BH}$ in low-mass galaxies.
        
        Observations indicate that the tight scaling relationships between $M_{\rm BH}$ and $M_{\rm star}$ (or velocity dispersion, $\sigma_\star$) found among massive galaxies tend to over-predict $M_{\rm BH}$ in low-mass galaxies \citep{Reines2015}. Using semi-analytic models, \citet{Pacucci2018} find that MBHs that are under-massive relative to the expected $M_{\rm BH} - \sigma_\star$ or $M_{\rm BH} - M_{\rm star}$ relations tend to have grown from weakly accreting low-mass seeds, which may fall below typical survey detection thresholds. This trend suggests that the observed $M_{\rm BH} - \sigma_\star$ (and similarly the $M_{\rm BH} - M_{\rm star}$) relation in massive galaxies only appears to extend to low-mass galaxies because those MBHs are detectable, when there may in fact be many undetected, low-luminosity MBHs that fall below the relation \citep{Baldassare2020}.
        
        Figure \ref{mbh-mstar-hist-luminosity} illustrates how varying the x-ray detection limit alters the detected distribution of $M_{\rm BH}\,/\,M_{\rm star}$ for dwarf galaxies at $z=0.05$. As a baseline, we show the underlying distribution of $M_{\rm BH}\,/\,M_{\rm star}$ without cuts on x-ray luminosity for dwarfs in each stellar mass bin. In each stellar mass bin, setting luminosity thresholds as low as $L_{\rm X}^{\rm AGN} > 10^{39}$ \ergs removes the majority of MBHs from detection, including many of the most under-massive MBHs. Setting a threshold at $L_{\rm X}^{\rm AGN} > 10^{39}$ \ergs misses $78\%$ of MBHs in dwarfs between $10^{8} < M_{\rm star} < 10^{8.5} M_\odot$, and $38\%$ of MBHs in dwarfs between $10^{9.5} < M_{\rm star} < 10^{10} M_\odot$. Increasing the threshold to $L_{\rm X}^{\rm AGN} > 10^{41.5}$ \ergs samples a strongly biased sample of MBHs, keeping only those few with $\log\left(M_{\rm BH}^{\rm acc}\,/\,M_{\rm star}\right) \sim -3$. However, the median of the distribution changes little regardless of detection threshold, different from what is found by \citet{Pacucci2018}.
        
        Figure \ref{mbh-mstar-hist-contamination} further shows how varying the acceptable contamination from XRBs and hot gas impacts the detected distribution of $M_{\rm BH}^{\rm acc}\,/\,M_{\rm star}$, where we define $f_{\rm cont} =  L_{\rm X}^{\rm AGN}\,/\,L_{\rm X}^{\rm XRB+Gas}$. Throughout this work we have adopted $f_{\rm cont} = 2$ as the standard contamination threshold. Setting a contamination threshold as low as $f_{\rm cont} = 1$ similarly misses the most undermassive MBHs. A threshold set at $f_{\rm cont} = 1$ misses $77\%$ of MBHs in dwarfs between $10^{8} < M_{\rm star} < 10^{8.5} M_\odot$, and $60\%$ of MBHs in dwarfs between $10^{9.5} < M_{\rm star} < 10^{10} M_\odot$. As with setting luminosity thresholds, the median of the distribution changes little when setting contamination thresholds.
        
        Another complicating factor in MBH detection is distance from the galaxy center. Both simulations \citep{Bellovary2019,Bellovary2021} and observations \citep{Reines2020,Mezcua2020,Greene2021} have found MBHs at large radii from the centers of dwarf galaxies. Galaxies in \Rom{} have also been found to frequently host off-center MBHs at radii $>2$ kpc from the halo center \citep{Ricarte2021a,Ricarte2021}. Off-center MBHs frequently exhibit low luminosities \citep{Mezcua2020} and low accretion efficiencies \citep{Ricarte2021}, implying that the detection of off-center MBHs is intrinsically tied to luminosity thresholds and contamination. When selecting our initial sample of MBH hosts, we search for the brightest MBH within $10$ kpc of the halo center, which includes the majority of off-center MBHs. At $z=2$, $20\%$ of dwarfs with MBHs have their brightest MBH outside of $2$ kpc. Over time, bright off-center MBHs become even less common as the percentage drops to $11\%$ by $z=0.05$. Regardless of redshift, approximately $98\%$ of MBHs in dwarf galaxies are found within $10$ kpc of the halo center.
        
        \begin{figure*}
            \plotone{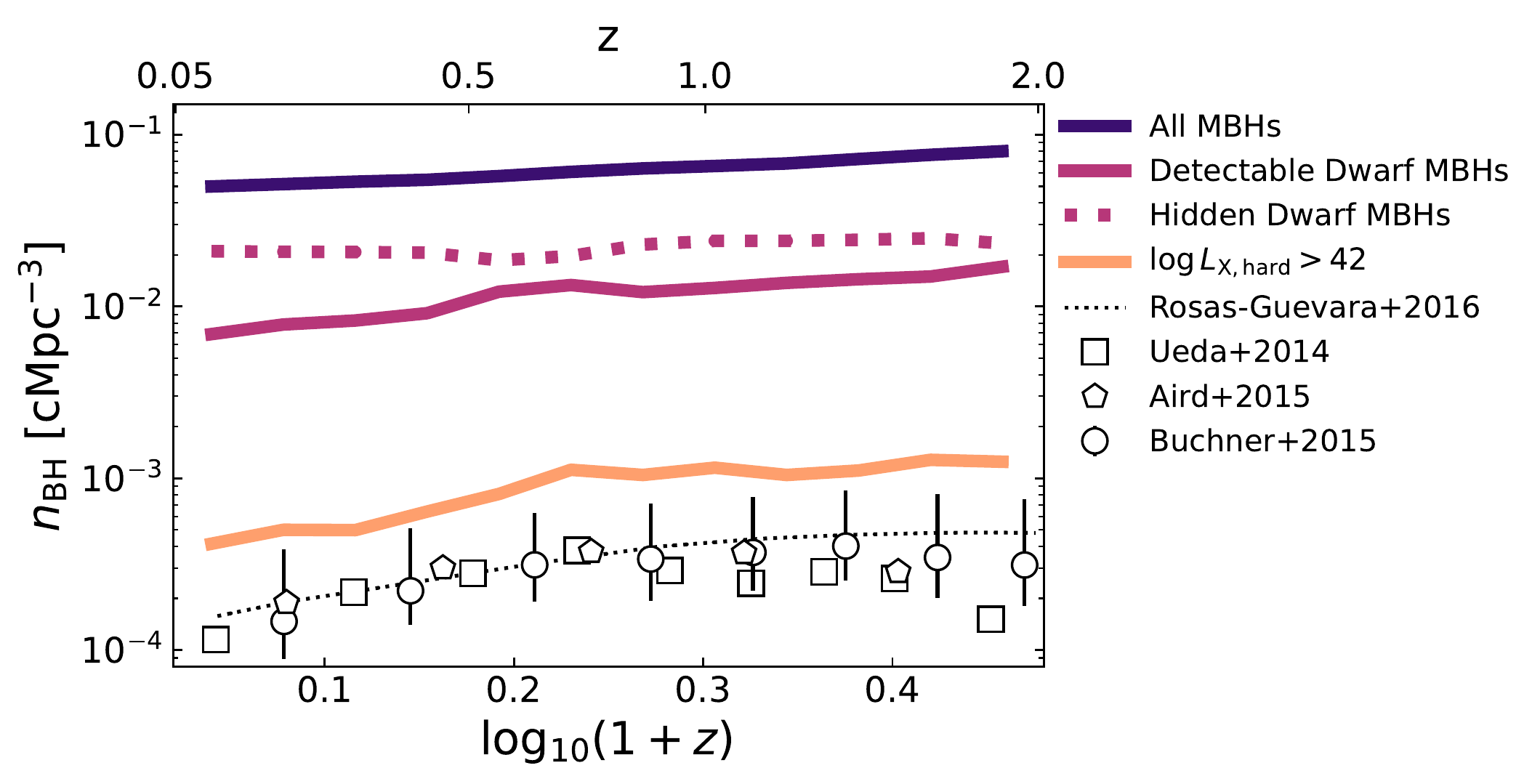}
            \caption{The spatial number density of the brightest MBHs within $10$ kpc of the halo center, versus redshift. Shown are all MBHs (purple); MBHs in dwarf galaxies with $L_{\rm X}^{\rm AGN} > 10^{39}$ \ergs and $L_{\rm X}^{\rm AGN} > 2L_{\rm X}^{XRB + Gas}$ (solid magenta); BHs in dwarf galaxies with $L_{\rm X}^{\rm AGN} < 10^{39}$ \ergs and/or $L_{\rm X}^{\rm AGN} < 2L_{\rm X}^{XRB + Gas}$ (dotted magenta); and all MBHs with hard $\left(2-10\right)$ keV luminosities $L_{\rm X, hard} > 10^{42}$ \ergs and $L_{\rm X}^{\rm AGN} > 2L_{\rm X}^{XRB + Gas}$ (solid gold). The number of hidden MBHs in dwarf galaxies outweigh detectable MBHs by a factor of $3$, and make up $40\%$ of the overall number of MBHs in \Rom{} at $z=0.05$. In \Rom{}, the number density of AGN with $L_{\rm X, hard} > 10^{42}$ \ergs is slightly higher than comparable estimates from observations \citep[][black square, pentagon, circle, respectively]{Ueda2014,Aird2015,Buchner2015}, and from the \textsc{EAGLE} cosmological simulation \citep[][black dotted]{Rosas-Guevara2016}.}
            \label{number-density-mbh}
        \end{figure*}
        
        To further quantify the combined impact of off-center MBHs, x-ray luminosity limits, and contamination on the detection of MBHs, Figure \ref{mbh-mstar-scatter} shows the $M_{\rm BH}^{\rm acc} - M_{\rm star}$ relation up to $M_{\rm star} < 10^{11} M_\odot$ with the same set of x-ray luminosity thresholds and acceptable contamination fractions. We distinguish between galaxies with central (within 2 kpc of the galaxy center) and off-center (outside 2 kpc of the galaxy center) MBHs. In cases where there are multiple central or multiple off-center MBHs, we choose the brightest one. For reference, we over-plot observed relations for massive galaxies from \citet{Schramm2013} and \citet{Kormendy2013}. 
        There exists a ``hidden'' population of MBHs at both low luminosities $\left(L_{\rm X}^{\rm AGN} < 10^{39}\,{\rm erg s}^{-1}\right)$ and high contamination $\left(L_{\rm X}^{\rm AGN} < L_{\rm X}^{\rm XRB+Gas}\right)$ that are undetectable by current instruments. Approximately $74\%$ of central MBHs in dwarf galaxies are hidden by these criteria. While the majority of hidden central MBHs lie directly along the observed relation, a significant number are found well below the relation. On the other hand, $88\%$ of off-center MBHs in dwarf galaxies are further hidden by low luminosities and/or contamination, and the majority are significantly undermassive. These results indicate that increasing the detection sensitivity alone would not allow hidden MBHs to be reliably detected since they are often heavily contaminated and sometimes exist at large distances from the galaxy center. Detecting these hidden MBHs with x-rays, especially MBHs that are far off-center, would require higher instrument sensitivity in addition to (i) a better understanding of XRBs + hot ISM emission in dwarf galaxies (e.g, in the low metallicity, low SFR regimes as discussed in \citealt{Lehmer2021}), and/or (ii) multi-wavelength imaging of the AGN.
        
        Although we predict that sensitivity limits will cause surveys to miss many undermassive MBHs, common detection limits only slightly impact the fit relation. Fits to the total (central + off-center) $M_{\rm BH}^{\rm acc} - M_{\rm star}$ relation below $M_{\rm star} < 10^{10} M_\odot$ at $z=0.05$ reveal that setting a $L_{\rm X}^{\rm AGN} > 10^{39}\,{\rm erg s}^{-1}$ detection limit simply shifts the relation up by $0.1$ dex while keeping the slope intact. The small change in predicted $M_{\rm BH}$ is not large enough to explain the $1$ dex difference found in \citet{Pacucci2018}. Their abundant under-massive MBHs are likely the result of their mixed seeding mechanism which generates both high-mass $\left(M_{\rm seed} \sim 10^{4} M_\odot\right)$ and weakly accreting low-mass $\left(M_{\rm seed} \sim 10^{2} M_\odot\right)$ seeds.
        
    \subsection{Number density of detectable MBHs} \label{number-density}
    
        Finally, we report on the spatial number density of hidden MBHs and AGN relative to the total number density of MBHs in \Rom{}. The initial number of MBHs within \Rom{} is set by the seeding prescription, while the evolution with time is determined by the merger rates. In particular, the seeding prescription determines where and with what frequency MBHs can form. Since the majority of MBHs in \Rom{} form prior to $z=5$ \citep{Tremmel2017}, the evolution of the number density from $z=2$ to $z=0.05$ is driven by MBH mergers. Although we do not directly report on MBH merger rates, it is worth noting that \citep{Volonteri2020} find MBH merger rates tend to increase among low-mass galaxies when moving to higher resolution simulations. Lower resolution simulations tend to miss mergers of low-mass galaxies, and hence underpredict the MBH merger fraction. It is likely that moving to higher resolutions than \Rom{} would lead to differing results for the evolution of the MBH number density.
        
        Figure \ref{number-density-mbh} tracks the comoving number density of MBHs as a function of redshift. We include lines for all MBHs in well-resolved halos, ``detectable'' MBHs in dwarf galaxies ($L_{\rm X}^{\rm AGN} > 10^{39}$ \ergs and $L_{\rm X}^{\rm AGN} > 2\,L_{\rm X}^{\rm XRB+Gas}$), and hidden MBHs in dwarf galaxies ($L_{\rm X}^{\rm AGN} < 10^{39}$ \ergs and/or $L_{\rm X}^{\rm AGN} < 2\,L_{\rm X}^{\rm XRB+Gas}$). We also show the evolution of the AGN number density from \Rom{} alongside x-ray observations \citep{Ueda2014,Aird2015,Buchner2015}, and \textsc{EAGLE} \citep{Rosas-Guevara2016} in the hard x-ray band with $L_{\rm X, hard} > 10^{42}$ \ergs.
        
        The total number density of MBHs decreases over time, from $n_{\rm BH} = 8\times10^{-2}$ cMpc$^{-3}$ at $z=2$ down to $n_{\rm BH} = 5 \times 10^{-2}$ cMpc$^{-3}$ at $z=0.05$. These number densities are within a factor of a few of estimates from \citet{Volonteri2010} who estimate $n_{\rm BH} = 0.02-0.1$ cMpc$^{-3}$ at $z=0$; slightly lower than those from \citet{Buchner2019}, who find $n_{\rm BH} \gtrsim 0.01$ cMpc$^{-3}$ at $z=0$ for their seed-independent empirical model; and in line with those from \textsc{EAGLE} \citep{Rosas-Guevara2016}, which finds $n_{\rm BH} = 7.2 \times 10^{-2}$ cMpc$^{-3}$ at $z=0.1$ with MBH seed mass $M_{\rm seed} = 1.48 \times 10^{5} M_\odot$. Among dwarf galaxies, visible and hidden MBHs are found in equal amounts at $z=2$, approximately with densities $2\times10^{-2}$ cMpc$^{-3}$, but at $z=0.05$ hidden dwarfs are more common by a factor of $3$, around $2\times10^{-2}$ cMpc$^{-3}$.
        
        AGN follow a slightly different shape and evolution than the full population of MBHs. The number densities of AGN are instead determined by the accretion history of the underlying population of MBHs. As explored in Section \ref{active-fraction}, AGN have decreased in both luminosity and prevalence since $z = 2$ within the simulation. The number of AGN above $L_{\rm X, hard} > 10^{42}$ \ergs peaks at $z=2$ around $n_{\rm AGN} = 1.2 \times 10^{-3}$ cMpc$^{-3}$. AGN are most rare at $z=0.05$, at $n_{\rm AGN} = 4.2 \times 10^{-4}$. These estimates are slightly higher than observations and \textsc{EAGLE} results in the same band \citep{Ueda2014,Aird2015,Buchner2015,Rosas-Guevara2016} at $z\sim0$. This difference is likely tied to our over-prediction of the AGN fraction.
                
\section{Caveats} \label{Caveats}

As discussed in \citet{Sharma2020}, the primary caveat of our analysis is the relatively large seed mass of MBHs in \Rom{}. Accretion rates in the lowest mass MBHs are inflated by the relatively large MBH seed mass. A lighter MBH seed mass might instead drive a different accretion history due to the lower average accretion rates. It is worth noting that our local dwarfs exhibit x-ray luminosities consistent with AGN observations down to $M_{\rm star} > 10^{8} M_\odot$ (see Section \ref{scaling-relations}), and follow the observed $M_{\rm BH} - M_{\rm star}$ at low redshift \citep{Sharma2020}. It is not clear if a lighter MBH seed would push luminosities below what is observed or alter their agreement with local scaling relations. On the other hand, less accretion may mitigate the higher dwarf AGN fractions found in \Rom{} (see Section \ref{active-fraction}).

A lighter MBH seed mass may also lead to more frequent seeding of MBHs, though constraints from simulations indicate that a lighter seed formation channel may ultimately lead to similar present-day MBH number densities \citep{Greene2020}. Occupation fractions are often thought to differ between seed formation channels, though \citet{Ricarte2018} find that light seeds may produce a large range of occupation fractions, with overlap with heavy seed mechanisms. Given these results, it is not obvious how MBH occupation would be impacted if \Rom{} had a lighter seed mass.

There are a number of uncertainties that we do not fully address when presenting results for the detected fraction of MBHs in dwarf galaxies. For one, we do not take into account uncertainties on the contaminant relations themselves. The observed relations that govern how we calculate $L_{\rm X}^{\rm XRB}$ and $L_{\rm X}^{\rm Gas}$ themselves have uncertainties that impact the contamination fraction. For example, \citet{Schirra2021} find that the XRB luminosity may vary by up to $1$ dex at $z=0$ (and by more at higher redshift) between different models of XRB emission. We also do not include the effects of AGN obscuration, which is still virtually unconstrained in how much it impacts low-luminosity AGN \citep[e.g,][]{Schirra2021}. We also do not account for differences in star formation properties relative to observed dwarfs. The stellar masses and SFRs of dwarf galaxies in the simulation, which factor into calculating $L_{\rm X}^{\rm XRB}$ and $L_{\rm X}^{\rm Gas}$, may differ from real dwarfs. Indeed, results derived from six large-scale cosmological simulations by \citet{Haidar2022} indicate that it is possible to qualitatively reproduce the observed $M_{\rm BH} - M_{\rm star}$ and SFR$-M_{\rm star}$ relations, but still yield different estimates of the detected fraction. A closer examination of the quiescent fractions in \Rom{} may help validate the MBH detected fractions shown here.

As found in Section \ref{active-fraction}, comparisons with observations are sensitive to the choice of accretion model, bolometric corrections, and AGN obscuration. Active fractions are much closer to what is observed when we calculate luminosities with two-mode accretion, but \Rom{} internally uses a single-mode model when implementing feedback. It is only our post-processing results which suggest that the single-mode thermal feedback in \Rom{} does not properly emulate what happens in reality. A feedback model that depends on accretion rate, as found in some other cosmological simulations \citep{Sijacki2015,Weinberger2017,Dubois2016,Dave2019} may better match reality. If it is true that the radiative efficiency decreases for low $f_{\rm Edd}$, it is still unclear what form the feedback (and hence the radiative efficiency) should take at low $f_{\rm Edd}$. A closer examination of the Eddington fraction distributions, full spectral energy distributions, and obscuration fraction among dwarf AGN will be required to constrain dwarf activity in the future.

\section{Conclusions} \label{Conclusion}

In this work we explore the characteristics of activity in massive black holes within the high resolution cosmological hydrodynamic simulation \Rom{}. We study the population statistics of MBHs in galaxies above $M_{\rm star} > 10^{8} M_\odot$, including the occupation and number densities of MBHs in galaxies out to $z=2$. We focus on investigating the properties of MBH activity in dwarf galaxies between $10^{8} M_\odot < M_{\rm star} < 10^{10} M_\odot$. In summary, we find that:

\begin{itemize}
    \item Figure \ref{occupation-fraction} shows that the MBH occupation fraction at $z=0.05$ drops below unity for galaxies below $M_{\rm star} < 2 \times 10^{10}$, in broad agreement with observed constraints from x-ray selected AGN \citep{Miller2015}, x-ray observations of late-type spiral galaxies \citep{Desroches2009,Greene2012}, dynamical MBH estimates \citep{Nguyen2019}, and variability-selected dwarf AGN \citep{Baldassare2020}.
    \item MBHs in dwarf galaxies around $z=0.05$ follow established scaling relations between $L_{\rm X}^{\rm AGN}$, $M_{\rm star}$, and SFR that have been observed in dwarf AGN \citep{Mezcua2018, Birchall2020}. Figure \ref{observability-plot} shows these relations hold for AGN relatively uncontaminated by XRBs and hot gas emission, but break down at low stellar masses for weakly accreting and/or strongly contaminated sources.
    \item Dwarf AGN are rare in \Rom{} but not as rare as expected from x-ray observations. The dwarf active fractions in Figure \ref{active-fraction-mstar} evolve strongly with time, peaking at $z=2$ and dropping steeply toward the present day. Despite the steep evolution, the active fractions at $z=0.05$ are slightly higher than observations of local dwarf AGN \citep{Mezcua2018,Birchall2020}.
    \item Changes in radiative efficiency and bolometric corrections can dramatically affect activity among MBHs in simulated dwarf galaxies, as illustrated in Figure \ref{active-fraction-diff}. Constraining these quantities will require observations of Eddington fraction distributions and spectral energy distributions among dwarf AGN.
    \item We predict a considerable population of both central and off-center MBHs at $z=0.05$ that are undetectable by current x-ray facilities. These MBHs often exhibit luminosities lower than current x-ray detection limits $\left(L_{\rm X}^{\rm AGN} < 10^{39}\,{\rm erg s}^{-1}\right)$ in addition to high x-ray contamination $\left(L_{\rm X}^{\rm AGN} < L_{\rm X}^{\rm XRB+Gas}\right)$. Figures \ref{mbh-mstar-hist-luminosity} and \ref{mbh-mstar-hist-contamination} indicate that this population of hidden MBHs does not significantly change the observed $M_{\rm BH} - M_{\rm star}$ relation for central MBHs, though off-center MBHs are nearly all undermassive. Figure \ref{mbh-mstar-scatter} shows that $74\%$ of central MBHs in dwarf galaxies and $88\%$ of off-center MBHs in dwarf galaxies are hidden by low luminosities and/or high contamination.
    \item Figure \ref{number-density-mbh} shows that the number density of MBHs in \Rom{} are consistent with direct collapse seeding estimates from empirical models \citep{Buchner2019}, analytic models \citep{Volonteri2010}, and the \textsc{EAGLE} cosmological simulation \citep{Rosas-Guevara2016}. As expected from elevated active fractions, AGN are somewhat more common than in x-ray observations \citep{Ueda2014,Aird2015,Buchner2015}.
\end{itemize}


   





Detecting MBHs in dwarf galaxies with X-ray observatories has been a challenging endeavor, due to the combination of low intrinsic luminosities, the possibility of being off-center, and high chances of contamination from X-ray binaries, background quasars, and low surface brightness hot gas. Our work here highlights how a non-negligible fraction of MBHs are ``hidden'' to most observations due to a combination of low accretion luminosities and blending in with background sources. To maximize detection, one requires an X-ray observatory with both high sensitivity (to find faint sources) and high angular resolution (to disentangle a potential AGN from other X-ray sources). At present, the \textit{Chandra X-ray Observatory} is the only instrument with sub-arcsecond spatial resolution. The purported angular resolution of the planned \textit{Athena} telescope is 5 arcseconds, which is not sufficient to localize an AGN candidate and separate it from other sources. Only a probe-class X-ray mission, such as that recommended by the Astro2020 Decadal Survey \citep{Astro2020} (and based on the design of the \textit{Lynx Observatory}, or as planned for the \textit{AXIS} observatory) can meet the needs for discovering these elusive objects.

In future work, we will further explore the potential connection between AGN activity and the star formation quiescence we find in dwarf galaxies.

\section{Acknowledgements}

RSS and AMB acknowledge support from the National Science Foundation under grant No. NSF-AST-1813871. MT is supported by an NSF Astronomy and Astrophysics Postdoctoral Fellowship under award AST-2001810. JMB acknowledges support from NSF AST-1812642 and the CUNY JFRASE award. \Rom{} is part of the Blue Waters sustained-petascale computing project, which is supported by the National Science Foundation (awards OCI-0725070 and ACI-1238993) and the state of Illinois. Blue Waters is a joint effort of the University of Illinois at Urbana-Champaign and its National Center for Supercomputing Applications.  \Rom{} used the Extreme Science and Engineering Discovery Environment (XSEDE), which is supported by National Science Foundation grant number ACI-1548562. RSS thanks Kristen Garofali for helpful conversations. We thank the anonymous referee for their insightful feedback that improved this manuscript.

\bibstyle{aasjournal}
\bibliography{paper}
\end{document}